# Macroscopic quantization of gravity

M. Y. Amin[1]
Faculty of engineering, Ain-Shms University, Cairo, Egypt

**Abstract**
The moon is receding from earth at an average rate of 3.8 cm/yr [6][7][9][12]. This anomaly cannot be attributed to the well-known tidal exchange of angular momentum between earth and moon [8]. A secular change in the astronomical unit AU is definitely a concern, it is reportedly increasing by about 15 cm/yr [9][10], in this letter; the concept of macroscopic quantization of gravity is introduced to account for these anomalies on theoretical basis. Interestingly, it was found useful in measuring the speed of gravity! And what is more interesting is the fact that the concept is based on solid well known classical physics with no modifications to any standard model. It was found that the speed of gravity $c_g$ is in the range $10^4 \times c < c_g < 10^5 \times c$.

**I Introduction**

Quantum mechanics is a microscopic phenomenon; it was originally developed to provide an explanation of the atom, especially that of hydrogen wich represents the simplest case. The sharp line spectra of light emitted when exciting an atom gave a definit proof of the quantum states of energy that an atomic electron can occupy. The quantum theory was then exteded to describe almost every aspect of the microscopic world from the atomic size down to Planck's length [1].

It was always assumed that "macroscopic" properties of "classic" systems are direct consequences of quantum behavior of its microscopic parts. we are going to extend quantum mechanics to include the macroscopic scale as well, showing that there are some quantum phenomena occurring at the scale of meters, kilometers and even light years!

---

[1] Electronic address: myhmamin@yahoo.com

## II Gravitational potential energy!

One of the oldest identified forms of energy is gravitational potential energy, in classical and quantum field theory, potentials have no physical meaning by themselves (cannot be measured), and only their gradients have. Up till now the origin of this energy is not clearly understood and definitely the energy balance about a massive object cannot be explained, for example, if a slowly moving asteroid (coming from an infinitely large distance) enters the effective gravitational field of a large planet, it will be accelerated towards that planet's center of mass and collides with its surface, we simply say that the asteroid's gravitational potential energy is converted into kinetic energy during acceleration then converted to heat and sound upon collision, no word is said about where did this potential energy come from? And of course the energy balance of the system {planet + asteroid} before and after this process is not satisfied!

If general relativity is correct and the source of gravitational potential energy is not a massive object (like a planet or a star for that matter) but rather the curvature of space fabric itself around that object (stress-energy tensor), then gravitational potential energy is a direct strong manifestation of vacuum energy (quantum zero-point energy) that we experience in our every day life, and gravitational waves should be as common as electromagnetic waves, but since we do not know how gravity actually works, we still can't devise a method of detecting such waves. On the other hand, if gravitational energy results from a direct interaction between two massive objects, then gravitational shielding should have been noticed by now (at least during solar eclipses). It is clear that the question is still open and probably will stay that way for a long time!
Dr. Jesse L. Greenstein of the California Institute of Technology wrote:

> *The detection of gravitational waves bears directly on the question of whether there is any such thing as a "gravitational field," which can act as an independent entity. This fundamental field hypothesis has been generally accepted without observational support. Such credulity among scientists occurs only in relation to the deepest and most fundamental hypotheses for which they lack the facility to think differently in a comparably detailed and consistent way. In the nineteenth century a similar attitude led to a general acceptance of the ether* [5]

This argument is raised to set the stage for even stranger results presented shortly and to emphasize the fact that: even we know nothing about how gravity actually works, we do know several fundamental mathematical laws describing gravity to very good accuracy, and increasing our accuracy in describing gravity is certainly a forward step in our understanding of it.

## III Macroscopic quantum gravity

Although the name implies quantum mechanics, MQG is purely a classical concept that modifies none of the existing theories, and by classical I mean prior to Relativity and Quantum Mechanics, it is based on only one assumption:

- *Gravity is not instantaneous; it rather acts on a finite speed $c_g$*

All physicists agree that gravity is not instantaneous, which implies an infinite propagation speed, the debate was always about how fast gravity acts? One of the earliest attempts to find out the speed of gravity was made by Laplace (in 1805), he concluded that the speed of gravitational attraction must be at least $7 \times 10^6$ times the speed of light $c$. Reacently, (Flandern 1998) concluded that the speed of gravitational attraction must be $\geq 2 \times 10^{10}$ $c$ [11], he also argued that General Relativity (GR) while vigorously claim a finite speed of gravity (i.e. $c_g = c$), reduces to Newtonian gravity with infinite propagation speed in the weak-field, low velocity limit, he literally stated that:

> In short, both GR and Newtonian gravity use infinite propagation speeds with aberration equal to zero. In Newton's laws, that fact is explicitly recognized even though aberration and delay terms do not appear because of an infinity in their denominator. In GR, much effort has been expended in disguising the continued absence of the same delay terms by including retardation effects in ways that are presently unobservable and ignoring aberration. Every physicist and physics student should be at least annoyed at having been tricked by this sleight of hand, and should demand that the neglect of aberration be clearly justified by those who propose to do so. [11]

It is clear that the main problem in specifying the speed of gravity $c_g$ is gravitational aberration, which presents a challenge to any theory of finite gravitational speed including GR. In fact it is aberration that is the key to a solution for this problem! In the following we will carefully study what happens during the period of **gravitational aberration**, which we consider a quantum value despite its macroscopic scale.

We will assume in the next analysis, for the sake of simplicity, that a much smaller body is orbiting a much larger one, so it is not a classical two-body problem in a sense, but rather a point mass in orbit around a massive object, then equations of motion could be modified for more complex systems about barycenter of mass and could be adapted for numerical solutions on computers. Assume a small spherical body of mass $m$ is orbiting a much larger spherical one of mass $M$ with the orbital radius $r$ (center to center) in a perfectly circular orbit, as shown in fig.1 below:

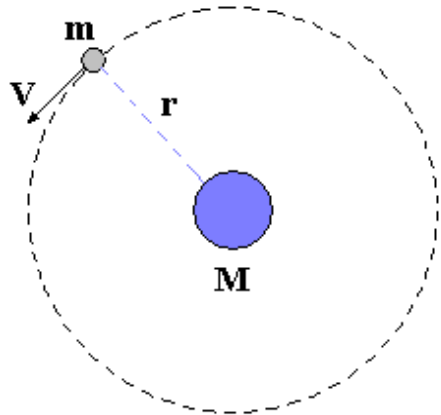

Fig. 1

The classical condition for a stable circular orbit is that the gravitational force on body m is iqual to the centripetal force needed to keep it in orbit:

$$\frac{mV^2}{r} = G\frac{mM}{r^2}$$

$$V = \sqrt{\frac{GM}{r}} \quad (1)$$

Where $G$ is the universal gravitational constant and $V$ is the instantaneous linear speed of the body $m$. Now according to our assumption of finite gravitational speed $c_g$ we introduce the concept of *minimum time of gravitational interaction* $\tau$ that is defined by:

$$\tau = \frac{r}{c_g} \quad (2)$$

This is the well known gravitational aberration time, but here we will call it the ***quantum time of interaction,*** that is the minimum time interval needed for two bodies (separated by $r$) to interact gravitationaly, regardless of the actual mechanizem of gravity, and of the actual speed of gravity (be it $c$ or $7\times10^6\times c$ or even $2\times10^{10}\times c$) this time interval has a physical meaning by itself and cannot simply be ignored!. During this interval $\tau$, there is no form of communications between the two bodies and if one body moves a distance (no matter how big or small) during that interval, the other one has no way of knowing where it is or how its velocity have changed!. This is a very good reason to consider the gravitational force acting on iether of them is constant (both magnitude and direction) during that period of time. We will analyze the motion of the small body m at each quantum time interval $\tau$ and see what happens there:

- The motion of body m has a linear component in the direction of it's instantaneous linear speed *V* and by the end of $\tau$ this linear component will have moved it a linear distance of *Δx* in the direction of *V* and if there were no gravity at all, a radial distance *Δr* as shown in fig.2.

- But the body also experiences a constant gravitational force perpendecular to its instantaneous linear velocity *V* vector that effectivly gives the body m a radial velocity component that points towards the center of mass and by the end of $\tau$ it will be pointing towards an imagenary point as shown in the figure, (the figure is exaggerated to illustrate the point, but in real systems like the earth-moon sytem these two centers practically coincide).

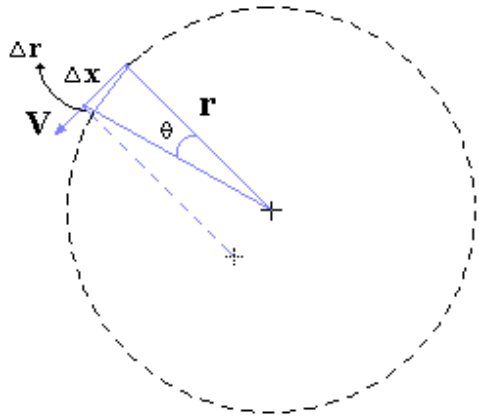

Fig. 2

We have :
$$\Delta x = \tau V$$

And:
$$\Delta r = \sqrt{r^2 + \Delta x^2} - r = \sqrt{r^2 + \tau^2 V^2} - r$$

Substituting for $\tau$ from (2) gives:

$$\Delta r = \sqrt{r^2 + \frac{r^2 V^2}{c_g^2}} - r \qquad (3)$$

- During the time interval $\tau$, the motion of body m is a combination of its linear motion (by its own inertia) and a net average radial velocity component $v_{av}$ (almost perpendicular to its linear speed) that is the vector sum of its previous radial component (from previous $\tau$ period) and a newly gained component from constant gravitational force during this interval, so that effectivly at the end of $\tau$, the body should have literally dropped a distance of $\Delta r$ towards the center with a velocity of $v_{av}$.

- During each interval $\tau$, the motion of body m is very similar to a horizontal projectile with a constant gravitational force perpendicular to its motion with the result of a small parabolic trajectory. In effect, the orbit will be composed of **quantum macroscopic steps** of nonlinear parabolic steps.

It is clear that the set of all radial gravitational velocity components $v_i$ in one complete orbit is a finite closed sequence of $n$ elements, and it is also obvious that:

$$\sum_{i=1}^{n} \vec{v_i} = 0 \qquad (4)$$

Because a gravitational radial velocity component is cancelled out after one half orbit (180°), we can conclude that:

> *Even in the absence of any perturbations, a perfectly circular orbit does not exist in nature, and even when initial conditions allow a perfectly circular orbit, the orbit eccentricity will not equal zero after one complete revolution, and it will continuously drift towards an elliptical orbit.*

Now in addition to (1) we introduce the second condition of orbit stability:

$$v_{av} \tau = \Delta r \qquad (5)$$

Substituting for $\tau$ and $\Delta r$ from (2), (3) gives:

$$\frac{v_{av} r}{c_g} = \sqrt{r^2 + \frac{r^2 V^2}{c_g^2}} - r$$

And we have:

$$v_{av} = \sqrt{c_g^2 + V^2} - c_g \qquad (6)$$

The importance of (6) is two folds, one is that we can establish a critical condition of orbit stability from which we can explain anomalies like that of the moon recession, and the other is that we can actually calculate $c_g$ to a good precision using data from any known stable orbit, calculating $v_{av}$ from gravitational force, and $V$ accordingly at each quantum period $\tau$ around a complete orbit (a finite sequence) taking into account vector velocity components at each point is an easy software task (although it takes a lot of computer time)[*]. I did run a few simulations on a hypothetical two-body system, with no perturbations of any kind, it was found that:

- When $c_g$ was set to any value $< 3 \times 10^{12}$ the orbit was unstable with a net radial velocity component pointing away from the center which leads eventually to the escape of the smaller body.

- When $c_g$ was set to any value $> 3 \times 10^{13}$ the effect was reversed with a net radial velocity component pointing towards the center and the eventual collapse of the system.

So it is assumed, as a first approximation that:

$$3 \times 10^{12} < c_g < 3 \times 10^{13} \text{ m/s}$$

For accurate determination of $c_g$, (to the least significant digit) it is important to run a wide rage of simulations on different real systems with very accurate data (which by the way is not available in public domains), this is a several months (or even several years) job on fast computers, and it is important that this task could be done as soon as possible in order to settle the issue of gravitational speed once and for all.

At this point, it is tempting to make a prediction that could prove or falsify this concept of macroscopic quantum effects:

> *Since our own earth is gravitationally bound in an orbit around the sun, and that orbit is composed of discrete nonlinear steps of period $\tau$, we should expect the entire planet to be vibrating in the radial direction (center at the sun) at a frequency of about $1/\tau$ and an amplitude of about $\Delta r$.*

---

[*] I already wrote a software program (MQGCALC) and it is available with source code for review.

This prediction could be easily checked out using seismology data; it is well known that seismologists can record even the slightest global vibrations of the planet, and if we take $c_g$ to be in the order of $10^4 \times c$ then $\tau$ will be 0.049866 sec and the expected frequency will be about 20 Hz with an amplitude of about 0.007 mm. This signal is well within the background noise signals recorded by seismographs every day. We can also predict a pattern for these vibrations alternating between traverse and longitudinal twice every 24 hours at a fixed point on the equator.

**IV Conclusion**

This concept needs to be tested and verified first before it could be useful in addressing the previously mentioned anomalies, and it might yield some new interesting macroscopic quantum behavior, for example: In analogy to microscopic quantum mechanics, planets orbits may show macroscopic quantum behavior in a sense similar to that of electrons around a nucleus, that is, a planet's orbit may be apparently unstable while in fact it is in a transition between two stable orbits (two macroscopic quantum states), this transition could be initiated by some sort of external perturbation, but the transition period is so long to be noticed in the case of planets and stars. A much simpler and faster domain for discovering these macroscopic quantum effects will be studying artificial satellites and their anomalies. Incorporating MQG into current models of astronomy may also help in the analysis of anomalies like that of the galaxies rotation, and even the pioneer anomaly!